\documentclass[11pt]{article}

\usepackage{graphicx}
\usepackage{amssymb}
\usepackage{latexsym}
 
\usepackage{amsmath,amsthm,epic}
\newcommand{\der}[2]{\frac{\mbox{d}#1}{\mbox{d}#2}}
\newcommand{\pder}[2]{\frac{\partial#1}{\partial#2}}

\def\mlr {Marc Lachi\`eze-Rey}
\def\frl{Friedmann-Lema\^ \i tre}

\def\rw{Robertson-Walker}

\def\spt {space-time}
\def\eg {{e.g.}}
\def\minks {Minkowski spacetime}
\def\Minks {Minkowski spacetime}
\def\calM {{\cal M}}
\def\calO {{\cal O}}
\def\calW {{\cal W}}
\def\diese {\sharp}
\def\bemol {\flat}
\def\imply {\Rightarrow}
\def\wrt  {w.r.t.}
\def\T{{\rm T}}

\def\d {\mbox{d}}

\def\gr {general relativity}

\def\spt {spacetime}

\def\coord {\mbox{coordinate}}

\def\lr{Lachi\`eze-Rey}

\def\calH {{\cal H}}

\def\ie {{i.e.}}

\def \guill {\textquotedblleft}
\begin{document}
\title{The covariance of GPS \coord s and frames}
\author{\mlr\\
CNRS APC,  UMR 7164\\
Service d'Astrophysique, CE  Saclay, \\
91191 Gif sur Yvette Cedex, France  \\
e-mail: marclr@cea.fr
}
\maketitle

\abstract{We explore, in the general relativistic context,  the  properties of  the recently introduced GPS \coord s, as well as those of the associated  frames and coframes that they  define. We show that they are  covariant, and completely independent of any observer.
We show that standard spectroscopic and astrometric observations allow  any observer    to measure, from  (i) the values of the GPS \coord s at his position, (ii) the components of his [four-]velocity   and (iii) the components of the metric in  the GPS frame. 
This  provides to this system an unique value   both for conceptual discussion (no  frame dependence) and for  practical use (involved  quantities are directly  measurable):  localisation, motion monitoring, astrometry, cosmography, tests of gravitation theories.

We  show explicitly, in the  general relativistic context,   how an observer may   estimate its position and  motion, and     reconstruct   the components of the metric. This arises  from  two main results:   the   extension of  the velocity fields of the probes to  the whole (curved) \spt;  and the identification  of the  components of the observer's velocity  in the GPS frame with the (inversed)  observed redshifts of the probes. Specific  cases (non relativistic  velocities;   Minkowski and \frl ~\spt s;       geodesic  motions) are studied in details. }
\date{}

\section{Introduction and notations}\label{prelim}

In \gr, \guill ~real " quantities are represented by covariant objects, \ie, defined  independently of any frame or system of \coord s.  
A moving object (hereafter, a \emph{probe})  is represented by its worldline, and the associated [four-]velocity, [four-]acceleration vectors... All these entities are   covariant.  An observer, as a particular case of  moving object, is also  represented   by its (covariant)  world line,   velocity and acceleration. 
An observation requires an observer and an observed system, both   described by covariant quantities. Any observable quantity is a scalar  covariant combination (usually a tensorial contraction) of covariant quantities associated to the observer and the observed system. 

Most often, calculations in \gr ~ cannot be performed without the auxiliary introduction of specific frames (or \coord s). This  perfectly justified procedure introduces  however a risk of  misinterpretation of the      intermediary frame-dependent (thus, non covariant) quantities   introduced. 
In addition, one has to distinguish between observer-dependent and observer-independent results.
Thus, one has usually the alternative between covariant and observable quantities on the one hand, and frame-dependent quantities on the other hand, which can be more easy to calculate but require  special care  to be  converted  into  really observable quantities. 

This motivates the introduction of the following GPS \coord s which are   covariant and observable. Let us emphasize that the quasi totality of \coord ~systems used in \gr ~calculations do not offer these advantages. 
In addition, this system of \coord s is completely observer independent. This allows easy comparison of measurements made by different observers, or by the same observer at different locations.
This makes possible  to know   what another observer would observe in the same situation. Finally, even for a well defined observer, there is no global  frame canonically defined (see, \eg,  \cite{Observer}), so that   any observer-dependent frame involves  an undesirable arbitrariness, which is removed  with the GPS \coord s. 

GPS \coord s have been  recently introduced  by Rovelli \cite{rovelli} and also, independently, by \cite{Coll} and \cite{Pozo} under the name \emph{emission \coord s}.
They  are covariant; they represent directly measurable quantities, and they are independent of the observer.  They show some similarities   with the \emph{optical}Ê \coord s introduced by \cite{Synge}, which however are  observer dependent.
The denomination \guill ~GPS~\coord s " comes from the fact that their definition assumes the existence of   four reference probes, a situation which corresponds to the well  known Global Positioning System (GPS, see for instance \cite{GPS}). 
However, their  use is much more general, since  they can  be defined from any system of \guill ~probes~",  that we define here as  objects  emitting a signal. They can be part  of  system of artificial satellites  (like  the GPS), but also  stars,   pulsars,   galaxies... This system of probes constitutes what is called a generic and (almost) immediate  \emph{location system} by \cite{Coll}. The only requirement is that an observer is able to measure 
the direction and the  arrival time of the emitted signal (in his proper time) and  its redshift. It is shown below how this allow to construct the GPS \coord s and different additional quantities. This makes GPS  \coord s   very convenient   for localization, astrometry, tests of gravitation theories, cosmography.
Note that the situation can be improved    in especially designed experiments, where the probes  send additional information, like for instance the instant  (in any   time \coord) of the signal emission. Moreover, a further improvement (\emph{auto-located} positioning systems,  \cite{Coll}) occurs  when  the probes also transmit  the values of the  proper time information that they receive from the other probes. But we will not consider explicitly such cases in this paper. 

In  \cite{rovelli}, Rovelli has introduced several fundamental properties of the GPS \coord s system, and derived their explicit calculation in \Minks ~(see also \cite{Coll} and \cite{Pozo}). Here we extend these calculations, and    introduce the \emph{frames} and \emph{coframes}  defined by  the GPS  \coord s. We  show that they are also  covariant, in the general relativistic meaning; that they do not depend on any observer, and that their components  can be measured, in a manner that we specify explicitly.    Our  calculations apply  to an arbitrary \spt ~with curvature. They are valid in the context of any metric theory (including \gr) and thus allow potential applications for  checking  gravitation theories.

{\bf Observer's frames}

On the other hand, it may be convenient for an  observer to  chose a preferred frame for making his calculation,    expressing and  interpreting  his  observational results.  The  timelike direction defined by his four velocity   allows a   local  (strictly speaking, infinitesimal)  space +  time splitting. 
We will  indicate precisely the passage  between an  
arbitrary  frame (for instance one linked to an observer)  and the GPS  frame.  We show   the  nice property that the (inverse) velocity components of the observer, in the GPS frame, identify with   the set of the redshifts he receives from   the probes.

After recalling some geometrical results of differential geometry (\ref{geom}), section (\ref{s1})   first considers the question of the observation of a probe by an observer, in a metric theory. Then, in (\ref{s3}), we study the \coord s, frames and coframes defined by a set of four such probes. In (\ref{applications}), we present potential applications, and finally (\ref{mink}), we apply our results to the simplified case of \minks, and recover some results already 
found in the literature.

\subsection{Geometrical preliminaries}\label{geom}

At any point $m$ of a differential manifold $\calM$ (here, \spt),  the tangent space $\T_m \calM$ is dual to the cotangent space $\T^*_m \calM$.
The duality is however  not canonical, in the sense that   there is no natural  association of  a one-form to a vector. There is however  a natural way to define the dual of a frame $(e_\mu)$ as the coframe $(e^\mu)$ defined through
$$<e^\mu,e_\nu>=\delta ^\mu_\nu.$$
Each $e_\mu$ is a vector; 
each $e^\mu$ is a one form, and the brackets represent the action of one-forms on vectors. Note that this frame duality does not depend on any metric.

We assume now a (Lorentzian)  metric $g$.   A \emph{null frame} $(e_\mu)$ is such that $$g _{\mu \mu}\equiv g(e_\mu,e_\mu)\equiv e_\mu\cdot e_\mu=0,~Ê\forall \mu$$ (no sum over indices).
      A \emph{null coframe} is such that $$g ^{\mu \mu}\equiv g(e^\mu,e^\mu)\equiv e^\mu\cdot e^\mu=0,~Ê\forall \mu.$$  
Note that the dual of a null coframe is not a null  frame, and reciprocally.

To any  vector $V$,  the \emph{musical isomorphism} (or \guill ~canonical isomorphism ") associates  the one-form $\bemol V$ defined through 
$$g(V,W) \equiv V\cdot W=<\bemol V,W>,~Ê\forall W.$$ Similarly, to a one-form $\theta$, one associates the vector $\diese \theta$ such that $\bemol (\diese \theta)=\theta$.

To  a  given frame $(e_\mu)$,   the musical isomorphism  associates   the   coframe  $(\bemol e_\mu)$. 
In general, it does not coincide with  the  dual coframe $(e^\mu)$.  Only when the  frame is orthogonal (as it  is common in standard  calculations),   the two  coframes coincide, up to constants. This is not the case  for  null frames or coframes.

In the following, we will consider the holonomic  coframe $(\omega ^\alpha=\d s^\alpha)$ associated to the GPS \coord s $(s^\alpha)$, and its dual  frame  
$(\omega _\alpha = \pder{}{s^\alpha})$; but also the musically transformed   frame    
$(\Omega _\alpha\equiv  \diese \omega ^\alpha)$, and its dual coframe 
$(\Omega ^\alpha \equiv \bemol \omega _\alpha)$.

Let us also recall that a frame is holonomic, when it is linked to (local) \coord s. In this case, one can write (locally):
 $$ \omega ^\alpha =\d s^\alpha,  ~Ê\omega _\alpha=\pder{}{s^\alpha} =\partial _\alpha.$$
 
\section{ Observing a probe}\label{s1}

\subsection{Probes}
Very generally, we call a \emph{probe} any object  emitting an observable signal,   which can be used as a reference. The only requirement is that an observer can measure the arrival time of a signal (in his proper time) and its redshift. 
As we will see below, this gives direct access to  the  proper time of the  signal emission  by the probe, which is at the basis of the formalism.
Typically, the probe  can be  a \emph{GPS satellite}, hence the denomination. It can also 
 be a \emph{star} (in particular a pulsar)  sending its radiation,    which may constitute an efficient astrometric  tool. This can be also  a distant  \emph{galaxy}, or a quasar, in cosmology.
In some circumstances,  the probe can be a member of an especially designed system of satellites (\eg, the GPS system or the LISA experiment). In this case, these satellites can send some additional information,  offering additional possibilities, including the tests of  metric theories. In  this paper, we do not explore  such opportunities, and only consider the basic case where we can measure the arrival time and direction, as well as  the redshift of the signals, from which results the (probe's) proper time of emission,~$s$.  

The motion of the probe (assumed to be given)  is defined  by its   time line $s \mapsto P(s)$, with $s$  its proper time. We do not assume  geodesic motion. 
 Although, in general,  the observer has no direct access to $s$, he can in fact  monitor its value  by integration of the redshift, as shown below. Thus we consider   $s$ as a measurable quantity.   
The (normalized)   velocity of the probe, $u\equiv \der{}{s} \mid_{P}$,  defines a vector field    along its world line. It verifies $u\cdot u=1$. 
We will show below how to extend this velocity field to the whole \spt. 

\begin{figure} 
\includegraphics[width=4in]{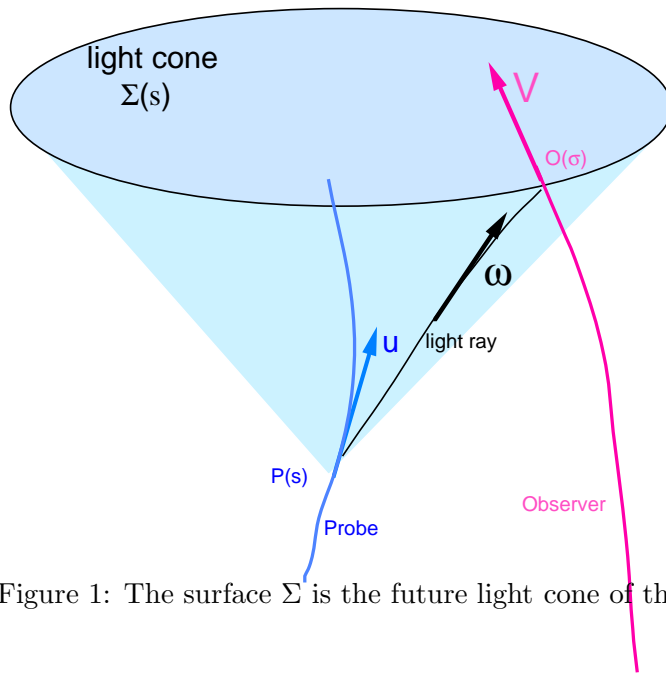}
\caption{The surface $\Sigma$ is the future light cone of the probe.}
\label{sigma}
   \end{figure}

The proper time of the probe, $s$, is defined along the world line $P$ only. We extend it as  a  function $x\mapsto s(x) $ defined  \guill ~everywhere " (hereafter, \guill ~everywhere " means the region of \spt ~where these \coord s are well defined; see the discussions by \cite{rovelli} and \cite{Pozo}) : at any  event $x$ in \spt, $s(x) $ is defined as  the proper time of the probe, when it emitted the light ray reaching $x$ (see the discussion in \cite{rovelli} for the case where many light-rays reach the observer).  In other words,  the hypersurface $\Sigma _{s_0}$, of equation $s=s_0$ is the future light cone of the probe at its proper time  $s_0$ (figure \ref{sigma}). This is a null hypersurface (for the properties of null hypersurfaces, see  \cite{gour}). 
It is a fundamental fact that $s$ is an observable quantity.

Each   $\Sigma _s$ admits   a one dimensional vector space of vectors orthogonal to it, at any point. These vectors are all proportional to $\diese \d s$.   Since the surface is null, these vectors  are also tangent to it, and they are null vectors ($\diese \d s\cdot \diese \d s=0$).  
Their  integral lines are null geodesics which constitute the null generators of $\Sigma _s$. They are  the future directed   light rays emanating from the event $P(s)$.
 In a null   surface, the null vectors   cannot be normalized like for a spatial surface. However,  one may select here   an unique orthogonal vector, from  the  form $\omega \equiv \d s$, which  is well defined over $\Sigma _s$ (except on the line $P$ itself, where  $\Sigma _s$ becomes singular). The canonical isomorphism (generated by the  metric $g$ of \spt)  transforms the one form  $\omega$ to the vector $ \Omega\equiv \diese \omega$ such that :
$$<\omega,  \Omega> \equiv g(  \Omega,  \Omega)\equiv   \Omega \cdot   \Omega\equiv g( \omega, \omega)=0 $$  
(see the detailed proofs in \cite{gour}). 
This implies  $$\Omega \cdot\nabla    \Omega \equiv \nabla_\Omega      \Omega =  \Omega \cdot   \nabla   \omega =0,$$ which means that $  \Omega $ and $\omega$ are parallelly transported by $  \Omega$,  that the null lines generated by $  \Omega$ are geodesic, with  $  \Omega$  an affine vector along them.  The  vector  $ \Omega $ is the \emph{frequency vector} \cite{bolos} of the light ray (here normalized so that the emitted frequency is unity).

Note that $  \Omega$ and $\omega$ are not defined on the world line $P$ itself. However, for each light-ray, it is possible to extend them by continuity, so  that they are parallelly transported : this allows us to consider   parallel transport along the null geodesics, including  their intersections with $P$.

Note that the function $s(x)$ obeys the equation
\begin{equation}
\label{world}
 \calW[P(s(x)),x]=0,\end{equation} where $\calW(x,y)$ is the \emph{world function} \cite{Synge}, defined as half the geodesic distance between the two points of \spt ~Ê$x$ and $y$.

\subsection{Extending the velocity of the probe}

Now let us define \guill ~everywhere " the  time like vector field $u$ obeying the following requirements:\begin{itemize}
  \item it is normalized:  $u \cdot u=1$;
  \item it is parallely transported along the null generators:$$     \nabla_\Omega  u= 0.$$ 
   Note that we have extended above the parallel transport up to the line $P(s)$ itself, and we require that $u$ is parallely transported in this way.
  \item It coincides with the velocity of the probe $P$ along its worldline. 
 Thus, it  may be seen as the velocity of the probe, transported everywhere by  the null generators.  
 \end{itemize}
Since $  \Omega$ is also parallely transported by itself, this implies  that $<\bemol u,  \Omega>=u\cdot  \Omega  $ is constant along the null generators (the light rays from $P$).  

\begin{figure} 
\includegraphics[width=4in]{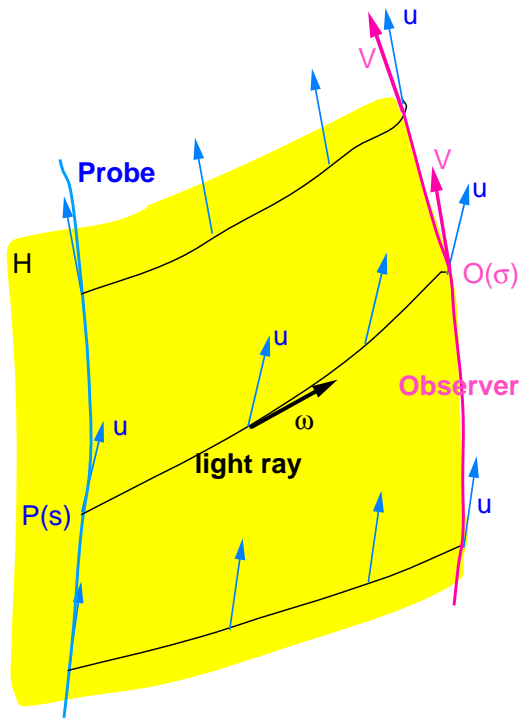}
\caption{The surface $\calH$ contains the world line of the probe and of the observers, as well as the light rays from the former to the latter.}
\label{hhh}
   \end{figure}

To apply    the third condition, we  consider the  2-surface $\calH$    generated by the probe world line, and  the light-rays reaching the observer, at successive moments (see figure \ref{hhh}). It also contains the observer's word line.  
The vector field   $u-V$ is well-defined on the observer's world line. It  represents the relative [four-]velocity of the probe with respect to the observer.

The two vector fields $u$ and $  \Omega$  are well defined on $\calH$,   including the world line~$P$ thanks to the mentioned extension. Moreover, $  \Omega$ remains tangent to  $\calH$. 
 On $P$,   the scalar product $  \Omega \cdot u=1$ because  there  the velocity $u=\der{}{s}$. Since this product is preserved by the null  generators,  as shown above, this implies that it keeps the value 1 on $\Sigma _s$: $$ \Omega \cdot u=1 \mbox{~everywhere},$$ one main result of this paper. 
This allows the   decomposition  $$  \Omega =u + \nu,$$ where the (spacelike) vector $\nu$ verifies $\nu\cdot \nu=-1$, $\nu\cdot u=0$.

\subsection{Observers and redshifts}

An  observer is  defined  by his worldline $\sigma \mapsto O(\sigma)$, along which flows his proper time  $\sigma$.  His   normalized  velocity $V\equiv \der{}{\sigma}$, with $V \cdot V=1$.
By definition, he sees the probe with a redshift 
  $$1+z=\der{\sigma}{s},$$ along his world line.

Since he can monitor this redshift as a function of his proper time,  $z(\sigma)$, 
this gives him access (up to an additional constant)  to the proper time of the probe at the moment of emission:
$$s(\sigma)=\int d\sigma~\der{s}{\sigma}=\int d\sigma~\frac{1}{1+z (\sigma)}+C^{te}.$$
This makes $s(\sigma)$ an observable quantity, as annouced.

On the other hand, it is easy to show that 
 \begin{equation}
\label{redd}
1+z=\frac{u(P)\cdot   \Omega (P)}{V\cdot   \Omega (O)} =\frac{1}{V\cdot  \Omega (O)}=\frac{1}{<\omega,V>(O)},
\end{equation} where $(O)$ and $(P)$ mean that the quantities are evaluated at observer's and probe's position respectively. Note that $z(\sigma)$ is a perfectly measurable, and covariant  quantity. But it  is observer dependent (it  depends on his velocity). Through this integral, the covariant quantity   $s(\sigma)$ is also measurable, up to a constant (we can imagine an optimal situation where this value is directly sent by the probe). The covariant \spt ~function $s(x)$ is  observer-independent.

  It is a fundamental fact   that the observer is able  to estimate   the value of $s$   by monitoring the redshift as a function of his proper time, $z(s)$, and by  integrating this relation. (In some specifically designed experiments, like in a future version of the GPS system, the probe may  emit  explicitly the value of $s$, which would allow additional possibilities.)  
 
By  projecting  the velocity $V$ of the observer onto the vector $\Omega$, one obtains
$$\Omega=\frac{1}{1+z}~Ê(V-n),~Ên\cdot n=-1,~n\cdot V=0,~V\cdot V=1.$$ The vector $n$ represents the spatial (unit) direction in which the observer sees the probe. 
For the observer, this vector is purely spatial in the sense that it is orthogonal to his velocity $V$, which defines the time direction for him.
Note that the observer is able to monitor this direction, which is also a covariant (but observer dependent)  quantity.

Taking into account $u\cdot  \Omega =1$, this relation implies
\begin{equation}
\label{zzz}
 1+z= u \cdot V-u\cdot n .\end{equation}
We emphasize that these relations hold in any \spt, for any observer, and for any metric theory.

\subsection{Non relativistic motion }

All formulae above hold for arbitrary velocities. If we work in the Solar System, or in the Galaxy (\ie, for instance, with pulsars), the  velocities involved are most often non relativistic, which corresponds to $z <<1$. All scalar quantities can be developed in this small parameter. We have for instance
$$  \Omega\cdot V=\frac{1}{1+z} \simeq 1-z~
\imply ~   \Omega \cdot (u-V)\approx z.$$
  This  relation was obtained in \minks ~by \cite{Tremaine}, and is at the basis of their analysis of pulsar timing. We have shown here that it remains true in an arbitrary \spt.

When  the probe (\eg, a pulsar)  has an intrinsic period $T$, the observer monitors  a period $$T_{obs}=T~(1+z)= 1-u\cdot n+\calO(z^2).$$
 
To continue, we may assume that the pulsar has an intrinsic variation $\dot{T}$ of its period. Then, by writing (\ref{zzz}) at two successive instants, one obtains the observed period variation
\begin{equation}
\label{TPOBS}
 \dot{T}_{obs}\equiv \der{{T}_{obs}}{\sigma}=\dot{T}~(1+z)+T~\der{z}{\sigma}.
\end{equation}
This is an exact formula. However, the  expression of $\der{z}{\sigma}$ is quite complicated.  \cite{Tremaine} have given an approximation of it, which is  valid for small redshifts, and in \minks.
 
 \subsection{Application to \Minks}\label{} 

The case of \minks ~is particularly simple. Without loss of generality, we can place the observer at the center of spatial \coord s: $O=(t=\sigma,0,0,0)$, so that  his velocity  has components $V=(1,0,0,0)$. 
A radial light-ray  arriving to the observer has   a (future directed)  affine tangent vector  $  \Omega$, with  components $ (K,-K,0,0)$, with $K>0$. And the probe has a velocity $(u^t,u^r,u^\theta,u^\varphi)$, with $u^r>0$ for a probe moving away from the source. Then it is trivial to form the scalar products $\Omega \cdot V=K$ and $\Omega \cdot u=K~(u^t+u^r)$, from which it results
$$1+z= u^t+u^r \approx 1+u^r,$$ where the latter approximation holds for non relativistic motion.

\subsection{Application to cosmology}\label{} 

In a \frl ~universe, with metric in the \rw ~form
$$g=dt^2-a(t)^2~[dr^2+f(r)^2~Ê(d\theta^2+\sin ^2\theta~Êd\varphi^2)],$$the observer is assumed at
the center of spatial \coord s, so that   his velocity  has components $V=(1,0,0,0)$. 
The null radial vector $\Omega =\Omega ^t~\partial _t -\frac{\Omega ^t}{a(t)}~Ê\partial _r $, with metric dual $\omega=  \Omega ^t~Ê\d t +a(t)~Ê\Omega ^t~Ê\d r$. The requirement to be an affine vector implies $\Omega ^t=K/a(t)$, with $K$ a constant  along the light ray. The value of the latter is fixed by the condition
$$1=u\cdot \Omega=\frac{K}{a(P)}~[u^t(P)+a(P)~u^r(P)],$$ where $(P)$ means evaluated at the position of the probe. 
This allows to find the expression for  the extension of the velocity field of the probe everywhere. In the case where the velocity is purely radial ($u^\theta =u^\varphi =0$), we obtain

$$u^t=\frac{1}{2} ~Ê(\frac{K}{a}+\frac{a}{K}),$$
$$u^r=\frac{1}{2a} ~Ê( \frac{a}{K}-\frac{K}{a}),$$
where we defined 
$$ K\equiv a_P~[u^t(P)-a_P~Êu^r(P)]=\frac{a_P}{u^t(P)+a_P~Êu^r(P)} $$

Using $\Omega\cdot V=\Omega^t=K/a(t)$,  we obtain the redshift 
$$1+z=\frac{a(t)}{K}= Ê\frac{a(t)}{a(P)}~\big[u^t(P)+a(P)~u^r(P)\big],$$
 which reduces to the usual formula 
 $1+z=  Ê\frac{a(t)}{a(P)} $
for a comoving galaxy, defined by $u^r(P)=0,~u^t(P)=1$.

\section{GPS coordinates and frames}\label{s3} 

\subsection{Coframes and frames}

Now we assume four probes $P^\alpha$ ($\alpha=1,2,3,4$) and   adopt, like in \cite{rovelli},   the four corresponding functions $s^\alpha$ as \coord s (see \cite{rovelli} and \cite{Pozo} for discussions about the range of validity of these \coord s, to which we   refer  as \guill everywhere "). As emphasized above, these \coord s are measurable.   They define  the coframe $(\omega^ \alpha \equiv \d s^\alpha)$.
By definition, the contravariant components of the metric tensor are given by $$g^{\alpha\beta}=
\omega^{\alpha} \cdot \omega^{\beta}.$$
As we have seen in (\ref{s1}), each $s^\alpha$ is a null function, so that
$$g^{\alpha\alpha}=
\omega^{\alpha} \cdot \omega^{\alpha}=0.$$
This means that $(\omega^{\alpha})$ is a null coframe.

\begin{figure} 
\includegraphics[width=4in]{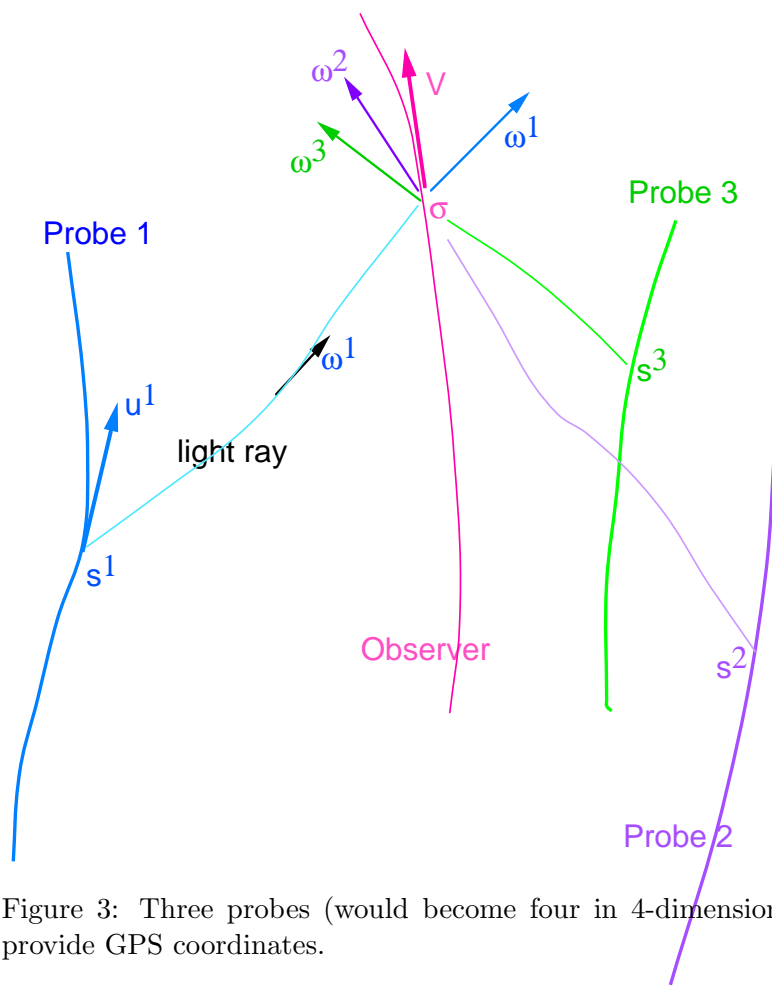}
\caption{Three probes (would become four in 4-dimensional \spt) provide GPS \coord s.}
\label{threeProbes}
   \end{figure}

This  coframe is holonomic. Its dual     frame,  given by      $(\omega_\alpha \equiv \partial_\alpha \equiv \pder{}{s_\alpha})$,  obeys   the duality relations
$<\omega^ \alpha,\partial_\beta>=\delta^ \alpha  _\beta$.
We have  by definition, $$g_{\alpha\beta}=\partial_\alpha  \cdot\partial_\beta.$$
The covariant components $g_{\alpha\beta} $ of the metric  are defined, as usual,  from the relations 
$g_{\alpha\beta}=g^{\gamma\delta}~g_{\alpha \gamma}~g_{ \beta\delta}$.
  Note that, in  general, $g_{\alpha\alpha}\ne 0$, so that $(\partial_\alpha)$ is not a null  frame

To the  system of \coord s $(s^\alpha)$,   are naturally associated the (holonomic) frame and coframes $ \partial_\alpha$ and  ${\omega}^\alpha$.
These frames involve only covariant quantities. They are defined \guill ~everywhere ".  They are completely independent of any observer. 

The musical isomorphism (see \ref{prelim}) defines 
 the vectors $\diese{\omega}^\alpha\equiv \Omega _\alpha$   (despite the upper index, the $\diese{\omega}^\alpha$ are vectors)  through  
 $$
 <\omega^ \alpha,   \Omega_ \beta>=\omega^ \alpha \cdot \omega^ \beta =  \Omega_ \alpha \cdot     \Omega_ \beta=g^ {\alpha \beta}.$$
 Since $g^{\alpha\alpha}=0$, $(  \Omega _\alpha)$ form a null frame.

 As vectors, the 
$    \Omega_\alpha$ 's  can be expanded in the $(\partial_\alpha)$ basis:
$$  \Omega_ \alpha= g^ {\alpha\beta}~  \partial _\beta \imply 
(  \Omega_\alpha)^\beta=(  \Omega_ \beta)^\alpha= g^ {\alpha\beta}.$$

See table 1 and figure \ref{threeProbes} for  illustrations of these frames and coframes.

\begin{table}
  \centering 
\label{Frames}
\begin{tabular}{|c|c|c|c|}
\hline
     $ \omega ^\alpha=\d s^\alpha$&$ \omega  _\alpha=\partial _\alpha$&$ \Omega ^\alpha=\bemol \omega  _\alpha=\bemol   \partial _\alpha $&$ \Omega _\alpha=\diese \omega  ^\alpha$  \\
\hline
  null & &   &null    \\
holonomic&holonomic&     &  \\
coframe&frame &   coframe  &   frame\\
\hline
\end{tabular}
\caption{Frames and coframes linked to the GPS \coord s} 
\end{table}

\subsection{Measurements}

Measurements are performed by an observer, that we assume given by his world line $\sigma \mapsto O(\sigma)$, $\sigma$ being his proper time, that he is able to read on his clock.  His  (normalized) velocity 
vector, $V \equiv \der{}{\sigma}$,  defines a  natural time direction for him,   at each point of his word line;  and, by othogonality, a set of  spacelike directions. In the following, we will assume that the observer used this   natural time+space splitting   to define  (locally) an ON  frame $(e_I)\equiv ( e_0 \equiv V,e_i)$, with $I=0,1,2,3;~ i= 1,2,3$. We do not care about the way the spatial part of this frame (the $e_i$'s)  is defined.

By spectroscopic observations,   the   observer is    able to estimate the redshifts corresponding to the signals emitted by the four probes, $ z^\alpha$.  As we will see, these observed quantities play a fundamental role.
As any vector, the observer's velocity   can be expanded in the basis as $V=V^\alpha~Ê\partial _\alpha$. From (\ref{redd}), it results    the very important relation 
$$V  \cdot  \Omega _\alpha =<\omega ^\alpha ,V>=V^\alpha \equiv \frac{1}{1+z^\alpha}.$$ This implies  $$V=\frac{1}{1+z^\alpha}~Ê\partial _\alpha:$$ the inverse  components of the observer's velocity reduce to the redshifts with respect to the probes. 
It results that, from the observed redshifts, the observer can derive the components of his velocity in the GPS frame, and thus monitor his motion.
Note that the four redshifts are linked by the relation $$g_{\alpha\beta}~Ê\frac{1}{1+z^\alpha}~Ê\frac{1}{1+z^\beta}=1.$$
 
The frames and coframes  defined by these \coord s are  also   covariant  and  independent of the observers. However, observations do not give direct access to the components of vectors,  forms or tensors. Those are generally evaluated in a given reference frame, chosen  for convenience (although the physics should be independent of such a choice). 

\subsection{The reconstruction of the metric}

All the  components of the metric tensor (in the GPS frame) identify with   covariant quantities  (in general,    a tensor is a covariant quantity, but not its components).  In addition, they are (as they must) completely observer independent.  Now we show that these components are directly  measurable.

The observer is able to project any quantity (like a vector or a tensor), at his position, parallelly and orthogonally to his velocity vector~$V$. He will interpret the projected quantities as temporal and spatial   components. Note that this decomposition, although of course observer dependent, remains perfectly covariant (does not require the introduction of any frame). In particular, 
$$   \Omega _\alpha=V^\alpha (V- n_\alpha );~V^\alpha\equiv \frac{1}{1+z^\alpha},$$ 
and its metric  dual (musical)  version$$  \omega  ^\alpha=V^\alpha ~ (\bemol V-\bemol  n_\alpha ),$$  where 
$$n_\alpha\cdot n_\alpha=
\bemol  n_\alpha\cdot\bemol   n_\alpha=-1,~V\cdot  n_\alpha=0,~ÊV\cdot V=1.$$
The    spatial normalized vectors    $n_\alpha$'s represent the (spatial)  directions of arrival of the signals of the four probes.
The   $\bemol n_\alpha$'s are the   spatial normalized one-forms, which represent the corresponding wavefronts.

The observer is able to  monitor these directions (in any frame), and thus to  estimate the scalar products $K_{\alpha\beta} \equiv   n_\alpha\cdot    n_\beta$ (obeying $K_{\alpha\alpha}=-1$). These measurable quantities are  frame independent.

Some easy algebra shows $$  n_\alpha =-V^\beta~ÊK_{\alpha\beta}~\partial _\beta \mbox{~(sum~on~}\beta).$$
Also, from the relations above, 
$$g^{\alpha\beta}=  \Omega _\alpha
\cdot   \Omega _\beta=\frac{1}{1+z^\alpha }~ \frac{1}{1+z^\beta} ~(1+K_{\alpha\beta}) $$ (no sum on indices). It results that all   the metric coefficients   $g^{\alpha\beta}$ are measurable by the observer. By   matrix inversion, he may also reconstitute  the matrix $g_{\alpha\beta}$.
This possibility to reconstruct the metric has also been considered in the 2 dimensional case by \cite{Coll}.

{\bf Observer dependent frames}

The observer may select three of these vectors, $\Omega _i$, that we label now with latin indices $i,j=1,2,3$.   With his velocity $V$, they constitute a perfectly valid frame, although observer-dependent. It  was introduced by Synge \cite{Synge} under the qualification of \guill ~Êoptical ".
Also,    the 
  three spatial vectors $n_i$, with   $V$,   constitute a frame  of a more usual form, with one timelike, and three spacelike vectors. In this case, we will label $V$ by the index $0$, and it is easy to establish that, in this frame, $$g_{00}=1,~g_{0i}=0,~g_{ij}=K^{ij}.$$ Thus, the observer is also perfectly able to reconstruct the metric coefficients in this frame.

\section{Potential applications}\label{applications}

Via the measured redshifts (or directly in appropriate situations), the observer  can monitor the   $s^\alpha$ as a function of his proper time. It is also able to measure the scalar products $K^{\alpha\beta}$, and thus to reconstruct completely the metric as indicated above.  Since the GPS frame is defined everywhere, it is possible  to perform the same measurements from other places : directly, following the observer's motion moving   in \spt; or indirectly, using measurements from other probes, which play the role of additional observers. The procedure above  provides  the metric  coefficients in the  different corresponding  points of \spt. By  derivations, this would give access to the components of the connection and of the curvature. 

In any case, the observer can monitor the redshifts along his world line, \ie, with respect to his proper time $\sigma$. Then, 
$$\der{z^\alpha}{\sigma}=A\cdot  \Omega_\alpha+V\cdot \der{  \omega_\alpha}{\sigma}.$$ This relation allows him to calculate his own  acceleration $A$. This  could be at the basis of experimental tests of gravitation theories.

If the probes send an  additional information, under the form of a variation of the period of a signal emitted by the probe (this is the case from a pulsar),   the   relation (\ref{TPOBS})  brings then  more constraints to the parameters of the system. They can be used for instance to determine the acceleration of the earth, or of the Solar System (see \cite{Tremaine}).

\subsection{Localization}

The observer may wish to use such \coord s for his localization. In some sense, this is a triviality   since the $s^\alpha$ are already  \emph{bona fide} \coord s, their measurements provide formally  a perfect localization.
In general, however, the observer is  not  interested by  localization with respect to the probes, but rather with respect to   a  more  usual system of \coord s $x^\mu$, like for instance a terrestrial one. In such a case, he would renounce to covariance, without  interest  for his purpose.

This requires no more than a  simple conversion   of variables $(s^\alpha)\leadsto (x^\mu)$. This possibility is guaranteed by  the fact that the $s^\alpha$ are true  \coord s. Explicitly, it requires    
the   explicit knowledge of the ephe\-me\-ris $P_{(\alpha)}^\mu(s^\alpha)$ of the four probes, in his favorite \coord ~system. For a system of artificial probes, they are perfectly known, and the conversion is simply a matter of calculations,  although there is no general analytic expression  for them in non  flat  \spt. In practice, the  ephemeris  has to be inserted in   the world equation.

This provides the formulae for the change of \coord s, $s^\alpha(x^\mu)$, and its inverse 
$x^\mu(s^\alpha )$. From this, the Jacobian provides the  transformation matrices for  the change of frames:
  $$~E^\alpha _\mu \equiv \pder{s^\alpha}{x^\mu} ,~ÊE_\alpha^\mu\equiv \pder{x^\mu}{s^\alpha}.$$

This allows us to express the vectors [one-forms] of this frame in the  other arbitrary frame [coframe].  $$\omega ^\alpha =\d s^\alpha =E^\alpha _\mu~Êe^\mu,~Ê\partial _\alpha =E_\alpha^\mu~Êe_\mu,$$Ê
$$e ^\mu =E_\alpha ^\mu~\omega^\alpha,~Êe _\mu =E^\alpha_\mu~  \partial_\alpha.$$Ê
Note that the last  formulas apply even when the new frame is not     holonomic (\ie, does not correspond to a system of \coord s). 

These formulae would be useful when an observer wishes to evaluate tensorial components in his personal frame (at the price of loosing covariance and observer's independence). Note that the knowledge of the explicit  correspondence requires to know the ephemeris of the probes.

\section{Application to  \minks}\label{mink}
 
In  flat  \minks,  the  properties of parallel transport imply   that the four velocities $u^
\alpha$ are vectors  constant on each light-cone. The positions $X \equiv O(\sigma)$ of the observer, and $P^\alpha$ of the probe may   be considered as   vectors, and we define the separation vectors (between the probes and the observer) $$D^\alpha \equiv P^\alpha-X.$$ Note that  $ D^\alpha\cdot D^\alpha=0$.

It is easy to establish the relations
$$\diese \d s^\alpha\equiv    \Omega_\alpha=\frac{D^\alpha}{D^\alpha\cdot u^\alpha},$$  
$$V^\alpha=1+z^\alpha=\frac{D^\alpha\cdot V}{D^\alpha\cdot u^\alpha},$$
$$\diese n^\alpha=V-\frac{ D^\alpha }{D^\alpha\cdot V}.$$

 It results $$g^{\alpha\beta}=\frac{D^\alpha \cdot D^\beta}{(D^\alpha \cdot u^\alpha)~( D^\beta  \cdot u^\beta)}.$$

If   the motions of the four probes are geodesic, we have 
 $P^\alpha=P^\alpha_0+u^\alpha~Ês^\alpha$, where 
  $P^\alpha_0$ represents  the origin  of the geodesic motion  of the probe $\alpha$.
Defining the four vectors $\xi^\alpha=X-P^\alpha_0$, we obtain an explicit expression for the GPS \coord s,  
 $$s^\alpha=\xi^\alpha \cdot u^\alpha-\sqrt{(\xi^\alpha \cdot u^\alpha)^2-\xi^\alpha\cdot \xi^\alpha}.$$
This expression  has been found  by  \cite{rovelli}, in the case where the four probes start at the same origin ($P^\alpha_0=0$).
 
  \subsection{Localization in \Minks}

The observer  monitors the GPS \coord s $(s^\alpha)$, and wishes to know the \coord s $x^\mu$ in his favorite frame, like the terrestrial one. The conversion requires to know the ephemeris of the four 
probes, $P^\mu_{(\alpha)}(t)$, where $t$ is an arbitrary parameter, that we take to be the time function in the  terrestrial frame.
Each  ephemeris may be easily explicited, along the worldline  of the probe,   as a function of its proper time, giving after conversion: $P^\mu_{(\alpha)}(t)=\Pi^\mu_{(\alpha)}(s^\alpha)$. Then, we have simply to solve the system of four equations
$$[X^0-\Pi _{(\alpha)}^0(s^\alpha)]^2= [X^1 - \Pi _{(\alpha)}^1(s^\alpha)] ^2+[X^2 - \Pi _{(\alpha)}^2(s^\alpha)] ^2+[X^3 - \Pi _{(\alpha)}^3(s^\alpha)] ^2,$$ for the four values of $\alpha$. Here, the
$\Pi _{(\alpha)}^\mu(s^\alpha)$ are the  functions derived from the ephemeris, and the unknown are the four $X^\mu$ which represent the \coord s of the observer in the desired frame.

Of course, the system becomes more complicated when there is curvature, but without problem to be solved numerically.

  \section{Conclusion and discussion}
    
    The GPS \coord s and corresponding frames offer a series of advantages. 
    First, they are covariant quantities. It 
   is not common in \gr ~calculations to dispose of \coord s or frames with this property.
   This provide to them an absolute and intrinsic character, which is however balanced by the fact that they depend on the choice of a set of probes. 
   
   Second, they are completely independent of the observer. This allows different observers, or an  observer at different locations in \spt,  to compare directly  different observational results and to interpret them,  without involving different \coord s systems and  frames. 
    
    We have shown how their use, with that of the associated frames and coframes, allow the complete reconstruction of the metric and, in some circumstances, of the curvature and connection.

    In the previous analysis, we have considered four probes and an observer. But the treatment of the  observer is  perfectly identical to those of the probes (the velocity $V$ of the observer is analog to that, $u$, of a probe; his proper time $\sigma$ is analog to that, $s$, of the probe). Thus the observer   can perfectly  be a probe of the same type, \ie, a fifth satellite.   
    Conversely, one of the probe can be chosen to be the terrestrial observer. In this case, one looses of course the  independence \wrt ~the observer, but this case   allows the easy  reconstruction of a  frame with more usual properties,  involving  timelike (for instance $V$) and spacelike vectors. 
    
    The possible applications are numerous.  This allows not only the  (relative) localization of the terrestrial observer, but also the monitoring  of its motion (velocity and acceleration), as well as the reconstruction of the metric, curvature and connection components.  
The latter possibilities open 
the opportunity  to combine the different relations established above  (which  remain valid in any   metric  theory, not only in \gr)   to provide   checks of the gravitation theories.  
    
    As we have remarked, the properties of a system of pulsars show many similarities with those of a GPS system. This allows to use such a system for precise astrometry, in the general relativistic context (see for instance the discussions in 
  \cite{Tremaine}).
  
  Finally, the situation also applies to cosmography.
  Although the metric of a \frl ~Êuniverse model is generally put under the \rw ~form, it is perfectly possible to express it by using GPS \coord s defined from a set of four distant galaxies, which may be assumed (or not) in free fall. This is feasible  in different manners, which correspond to different choices of the probes. 
  
  More generally, \emph{any metric} can be expressed in a GPS frame. This can be done     directly, by searching for the coefficients expressing the   change of frames. More easily, this could result from the choice of a set of convenient  probes, from which the \coord s and frames are defined as above. 
  
  The exploration of such possibilities, for systems of artificial satellites, for cosmology, and for the Schwarzschild metric, are in progress.
  
  \noindent
\textbf{Acknowledgments.} 
I thank B. Coll and  J.M. Pozo for useful comments on a first version of this paper.
 

\end{document}